\documentclass[aps,twocolumn,showpacs]{revtex4}

\begin{document}
\title{Non-universal Interspecific Allometric Scaling of Metabolism}

\author{Jafferson K. L. da Silva$^{1,\ast}$ and Lauro A. Barbosa$^{2,\dagger}$}

\affiliation{
$^{1}$ Departamento de F\'\i sica, Instituto de Ci\^encias Exatas, Universidade Federal de Minas Gerais\\
C. P. 702, 30123-970, Belo Horizonte, MG, Brazil\\
$^{2}$ Instituto de F\'\i sica de S\~ao Carlos, Universidade de S\~ao Paulo\\
Caixa Postal 369, 13560-970, S\~ao Carlos, Brazil\\}

\date{\today}\widetext


\begin{abstract}

We extend a previously  theory for the interspecific allometric scaling developed 
in a $d+1$-dimensional space of metabolic states. The time, which is characteristic of all biological  
processes, is included as an extra dimension to $d$ biological lengths. 
 The different metabolic rates, such as basal (BMR) and maximum 
(MMR),  are described  by supposing that the biological lengths and  
time are related by different transport processes of energy and mass.
 We consider that 
the metabolic rates of animals are controlled by three main transport processes: convection, 
diffusion and anomalous diffusion.  Different transport 
mechanisms are related to  different metabolic states, with its own values for allometric 
exponents. In $d=3$, we obtain that the exponent $b$ of BMR is $b=0.71$, and that the aerobic sustained MMR 
upper value of the exponent is $b=0.86$ (best empirical values for mammals: $b=0.69(2)$ and $b=0.87(3)$). 
The $3/4$-law appears as an upper limit of BMR. The MMR scaling in different conditions, other 
exponents related to BMR and MMR, and the metabolism of unicellular organisms are also discussed.

\end{abstract}

\maketitle

\section{\label{intro}Introduction}

Several biological quantities change with organism size according to  particular rules 
\cite{peters-83,calder-84,schmidt-84}. 
It is common to believe that these rules are related with the euclidean geometry. 
However, in many cases the geometric pattern is not observed because physical constraints also 
limit how much an organism can be modified to cope with changes in scaling. Recently, considerable 
effort has been invested to understand the scaling of some of these variables under certain physical and 
geometrical constraints: the dimensions of long bones \cite{garcia-04,garcia-06},
 the basal metabolic rate (BMR) \cite{west-97,banavar-99,banavar-02,darveau-02,demetrius-06,dasilva-07}
 and the maximum metabolic rate (MMR) \cite{dasilva-07,taylor-81,barbosa-06}.
In this paper we are interested in the scaling of metabolic rate, which is the most studied variable  in 
 traditional allometry.

It is accepted and empirically tested that the metabolic rate $B$ and the body mass $M$ of almost 
all organisms are connected by a power law relationship $B=aM^b$, where $a$ is a constant and $b$ is 
the scaling exponent \cite{peters-83,calder-84,schmidt-84,kleiber-32}. 
The origin and the universality of the scaling exponent of metabolic rates is a 
subject of great controversies and there are several debates in the literature 
\cite{dodds-01,savage-04,kozlowski-04,white-05,glazier-05,makarieva-05,dasilva-06}.
In a recent paper \cite{dasilva-07}, we and a colleague  
proposed an unified theory for the interspecific allometric scaling of metabolism.
It was developed in a $d+1$ dimensional space of metabolic states of organisms
($d$ biological lengths and a physiological time). It is natural to 
include explicitly an extra temporal dimension in the analysis of allometric scaling because 
all biological process are time dependent. Moreover, in some cases this approach has produced a simple 
explanation for the problem with satisfactory results \cite{dasilva-07,blum-77,west-99,Ginzburg-08}.
 In that paper \cite{dasilva-07}, the authors supposed that each metabolic 
rate of organisms is characterized manly by one of two transport processes, namely, 
convection and diffusion. In this paper we consider the general case in which a metabolic rate of 
$3$-dimensional organisms can be characterized 
by one, two or  three of the transport processes: convection, diffusion and anomalous diffusion.
It is well known, that the transport in large distances is done by convection and the transport 
in small distances is done by  diffusion. A classical example is the oxygen transported from the heart until 
the capillaries by convection 
and from the capillaries to the cell by diffusion. However, the mechanism of transport of large molecules
inside a cell and between the cells of a tissue is still unknown and in many cases 
is suggested  to be an anomalous diffusion. The three kinds of transport implies also that
 we must now  deal with different characteristic times. 
 But they are all related if  the network delivery is optimal.
 
This work is organized as follows. We discuss the hypotheses of da Silva, Barbosa and Silva (SBS) 
\cite{dasilva-07} and present a new one,  and derive the main
equations in Sec. \ref{sec1}. In Sec. \ref{sec2} we rederive the scenarios for BMR of SBS 
work in our present context as limiting cases.  
The BMR of mammals and birds is studied in Sec. \ref{sec4} and the scaling of capillaries and of aorta is 
obtained in Sec. \ref{sec5}. The approach to describe the MMR of endotherms is presented in Sec. \ref{sec6} 
and the exponents of aorta and capillaries in the MMR conditions are discussed in Sec. \ref{sec7}. 
The metabolism of unicellular organisms is discussed in Sec. \ref{sec3}. 
We summarize our results in the last section.

\section{\label{sec1}Hypotheses and main relations}

Following SBS \cite{dasilva-07} we  use the mass density 
$\rho_{d+1}(L_1,L_2,\ldots , L_d,\tau)$ (mass per unit volume and unit time) 
and the  energy density $\sigma_{d+1}(L_1,L_2,\ldots , L_d,\tau)$ (available energy 
per unit volume and unit time) to characterize the metabolic state of organisms. 
The use of energy density is 
justified because in the metabolic processes ATP cannot be supplied from 
outside but must be synthesized within the organism (within the cells). The efficient 
use of substrates by  cells depends on the presence of an adequate quantity of mitochondria 
as power house \cite{porter-93} and secondly on adequate supply of fuels and oxygen. The fuels, which are 
directly related to the available energy $E$, are contained inside the organism 
but the oxygen flux is supplied by outside the organism. So, energy content is important 
to characterize the state of an organism. This can  also be illustrated with the MMR situation. 
The animal runs until it has no more available energy. Then it fall exhausted.
  
The first and second hypotheses are that  natural selection enforces the constraints of {\sl
scaling-invariant} (independent of body mass) $\rho_{d+1}$  and $\sigma_{d+1}$, during evolution. 
 The third hypothesis is that the scaling
of the metabolic states is determined by the {\sl dominant dynamical transport
processes of nutrients}. Moreover, these processes are characterized by  scaling-invariant
quantities (diffusion coefficient, average velocity, etc.).

Although we have $d$ biological lengths, each one with its characteristic time
$t_i$ ($i=1,2\ldots,d)$, we suppose that only one time $\tau$ is relevant to describe
the metabolic states. It means that the resources rates of all these
processes must be matched ($(1/t_1)\propto (1/t_2)\ldots (1/t_d)\propto (1/\tau)$)
(symmorphosis principle proposed by Taylor and Weibel \cite{taylor-81}).
Therefore we are considering optimal transportation networks (the new fourth hypothesis).

 It follows from the second hypothesis that $E=\sigma_{d+1}~\tau V_d$. Here, we have that
  $\tau V_d$ is the $(d+1)$-volume and $V_d=L_1L_2\ldots L_d$.
 Using the power definition ($P=dE/dt$), the energy can be written in terms of the metabolic rate $B$,
 the power averaged over the time scale $\tau$, as $E=B\tau$. 
 Therefore from the first and second hypotheses  we obtain an equation for
 the organism's mass, namely
\begin{equation}
 M=\rho_{d+1}~\tau V_d~,\label{eq1}
 \end{equation}
 and the following expression for the metabolic rate
 \begin{equation}
 B=\sigma_{d+1}~ V_d~~.\label{eq2}
\end{equation}
Note that Eqs. (\ref{eq1}) and (\ref{eq2}) are valid  for all  metabolic regimes.
 Different metabolic scaling regimes will appear  because there are different ways to transport
 nutrients.

\section{\label{sec2}Limit scenarios for the basal metabolic scaling}

\subsection{The BMR-3 scenario}

We discuss in this section some limit cases for the basal metabolic rate scaling.
Let us first study the BMR-3 scenario, a lower bound for all
metabolic scaling.
 We suppose that  all the transportation occurs  via diffusion, implying that
\[
  L_i=D_it_i^{1/2}~,~{\rm with}~~ i=1,~2\ldots,~d~,
\]
 where $D_i^2$ are the
 scaling-invariant diffusion coefficients and $t_i$ are characteristic times.
  Since the  resource supply rates must be matched ($1/t_1\propto 1/t_2\ldots 1/t_d\propto 1/\tau$),
  we have only one time scale ($\tau$) and only one relevant length, namely
\[
L_1\propto L_2\propto\ldots\propto L_d\propto L~.
\]
 Note that the biological volume is given by $V_d\propto L^d$.
 Since $L_i=D_i \tau^{1/2}$,  we obtain from Eq. (\ref{eq1}) that
$\tau\propto M^{2/(d+2)}$. This relation furnishes how $L$ depends on $M$, namely
$L\propto M^{1/(d+2)}$, and
we can use Eq. (\ref{eq2}) to obtain that
\begin{eqnarray*}
L &\propto& M^\frac{1}{2+d}~,\\
\tau &\propto& M^\frac{2}{2+d}~,\\
B &\propto&  M^\frac{d}{2+d}~~.
\end{eqnarray*}
 In $d=3$, the metabolic exponent is $b=3/5$. Since these  transportation processes
 are the slowest ones, this value is a lower bound for the exponent $b$ for
 all metabolic situations.
 Note that this scenario can, in principle, describe the
  metabolic rate of very small organisms because diffusion over short distances is fast.

\subsection{The BMR-2 scenario}

For larger organisms, transport by convection is utilized on large length scales because diffusion is slow. 
 In the  cardiovascular system  of mammals, for example, blood circulates in a ballistic regime until the
 capillaries, where diffusion play the main role. Therefore
  we consider first that the BMR is  driven by
 {\sl ballistic transport}, namely
 \[
 L=v_0\tau~,
 \]
 where the velocity $v_0$ is scaling-invariant.
 Then we must taken in account the other metabolic steps.
   In a ``cylindrical'' symmetry
  we have $L_1\propto L=v_0t_1$ (ballistic term) and $d-1$ lengths  $R_i=D_i t_i^{1/2}$ 
  (diffusion terms). $v_0$ and  all $D_i$ ($i=2,3,\dots,d$) are scaling-invariant. Since the delivery of the
  network is optimal ($t_1\propto t_2\ldots\propto\tau$), it follows that $R_i\propto D_i \tau^{1/2}$. 
  From Eq. (\ref{eq1})
  we obtain that $\tau\propto M^{2/(3+d)}$, implying that $L_1\propto M^{2/(3+d)}$ and
  $R_i\propto  M^{1/(3+d)}$. Since the biological volume is $V_d\propto R^{d-1}L_1$, we
  obtain from Eq. (\ref{eq2}) that
\begin{eqnarray*}
L_1 &\propto& \tau\propto M^\frac{2}{3+d}~~,\\
R&\propto& M^\frac{1}{3+d}~~,\\
B &\propto&  M^\frac{1+d}{3+d}~~.
\end{eqnarray*}
 Then in $d=3$, we obtain  the $2/3$ law. Note that this result was obtained
  without mention of the area/volume ratio.

\subsection{The BMR-1 scenario}

 In this scenario all metabolic relevant lengths
are related to the ballistic transport, namely $L_i=v_it_i$ ($i=1,\ldots,d$).
 Using that all characteristic times are proportional to $\tau$ (fourth hypothesis), we find
 that $L_i\propto \tau$.
  In other words, there is only {\sl a single
 metabolic relevant length} $L\propto v_0\tau$ and {\sl a single time} $\tau$, both
  related to the ballistic transport. This scenario represents an upper bound for BMR.
 Since that $V_d\propto L^d$, we find from Eq. (\ref{eq2})that
\begin{eqnarray*}
L &\propto& \tau\propto M^\frac{1}{d+1}~~,\\
B &\propto&  M^\frac{d}{d+1}~~.
\end{eqnarray*}
 Therefore we find the $3/4$-law for $d=3$, namely $B\propto M^{3/4}$.
 This upper bound value is the same as these of West, Brown and Enquist \cite{west-97}  and 
 Banavar et al. \cite{banavar-99,banavar-02}.
  It is worth mentioning that Demetrius \cite{demetrius-06}
 found that the $b$ exponent of BMR should be in the interval $[2/3,3/4]$. His work is based in the
 integration of the chemiosmotic theory of energy transduction with the methods of quantum statistics.

 \section{\label{sec4}BMR of mammals and birds}

 From now on we use $d=3$ because cells and organisms are three dimensional objects. We also use
 a single time ($\tau$) for all transportation processes because all characteristic times are proportional
  to it.
 In order to study the BMR scaling of mammals and birds, let us discuss the nutrient transport in
 eukaryotic cells and between cells.
 The first biological length $L_1$ is related to the transport of oxygen and small molecules by
diffusion, namely
\[
L_1=D\tau^{1/2}~.
\]
 On the other hand,  large molecules can also be trapped  in  vesicles by
 macropinocytoses and pinocytoses and transported in direction of the nucleus.
  Note that a vesicle can carry a  relatively large quantity of fuel.
 Although the exact description  of vesicular transport is still unknown, we suppose it as
 an  anomalous diffusion process \cite{diffusion,alaor}, namely
\[
 L_2= D_x\tau^{(1/2)+x}~~.
\]
 The normal diffusion and ballistic transport processes occur when $x=0$ and $x=1/2$, respectively.
 This description is supported by works \cite{caspi-00,caspi-01} about the
 movement of engulfed particles on  eukaryotic cells.
 Beads placed on the peripheral lamella  of giant human fibroblasts are engulfed
 into the cytoplasm and move in direction of the nuclear region. In the lamella region
 the beads move ballistically with an average  velocity of $v\approx 1~\mu m/min$ ($L\propto vt$).
In the perinuclear region they move randomly within a restricted space and
 the authors have determined that $L\propto t^{3/4}$.
 Moreover, in a recent work of Neto and Mesquita \cite{neto-07} of optical microscopy, 
 the authors conclude that the movement of
 a macro pinosome inside a macrophage is ballistic ($L\propto v t$). The average velocity $v$ varies from
 $0.5~\mu /min$ to $2.0~\mu m/min$ depending on the radius of the macro pinosome.
  Therefore is quite probable that $x \approx 1/2$. We emphasize that a fuel vesicle is transported
  not only inside a cell but also from one cell to other one of the tissue.

 In the case of the BMR of mammals and birds,
 there is also a biological length
\[
 L_3=v_0\tau
\]
 related to the transport by convection utilized on large length scales.
  For example, we find in mammals the cardiovascular system that
 transports blood to the capillaries.

  To obtain the exponent $b$ we first evaluate $V_3$ in terms of $\tau$, namely
  $V_3\propto DD_xv_0\tau^{2+x}$. Then, we use the relation between mass and $\tau$
  (Eq. \ref{eq1}) to find how $\tau$ depends on $M$ ($\tau\propto M^{1/(3+x}$). Finally,
  we use  Eq. (\ref{eq2}) to obtain how $B$ depends on $M$. It follows that
  \begin{eqnarray*}
  B&\propto& M^{\frac{2+x}{3+x}}~,\\
  \tau&\propto&  M^{\frac{1}{3+x}}~.
  \end{eqnarray*}
  The case $x=0$  give-us $b=2/3$ and correspond to the BMR-2 scenario,
 where we have two lengths related to diffusion and one related to convection. This scenario
 yields the $2/3$ law without mention of the area/volume ratio.
 The upper limit for
  the BMR, the BMR-3 scenario, is obtained when the three lengths are related to convection.
  We obtain in this case the $3/4$-law.
   When $x=1/2$
 the vesicular transport within a cell is ballistic and we have that $b=5/7\approx 0.714$. Since it is
 quite probable that $x\approx 1/2$, the BMR exponent of mammals and birds is close to  $b=5/7$.

The empirical and predicted values of the BMR exponent $b$ of mammals and birds
 are shown in Tab. \ref{tab1}.
  This is the most analyzed and discussed allometric scaling in the last years.
  Note that all empirical
  values for $b$ are in the predicted range, except the one ($0.737(26)$) obtained by
  Savage et al. \cite{savage-04} with a data ``binning'' procedure. In such procedure
  the log-transformed data were averaged into equally spaced data points in order to
   achieve equal weight to all body size intervals and prevents phylogenetic relatedness.
   However, the error bars do not exclude the
  upper value of the range $0.714$. Since their procedure has been criticized by Glazier \cite{glazier-05},
  we note that the same data set without the binning procedure furnishes
  $b=0.712(13)$, a value in good agreement with our prediction. Perhaps the more rigorous and complete
  study of the mammalian BMR exponent  is the work of White and Seymour \cite{white-05}(see also
  \cite{white-03}).
  The authors excluded large herbivores of the data due to their long fast duration
  required to reach the postabsorptive state of BMR and obtained $b=0.686(14)$.
  They note that such animals are typically fasted for less than $72~h$ before the measurement of
  $O_2$ consumption, while the postabsorptive state of ruminants may require 7 days to be reached.
  We think that large mammals must be included in the data. Perhaps the BMR value can be obtained
  by a time extrapolation procedure, in which some measurements are realized periodically after
  the initial fasting. It is quite possible that this procedure will slightly raise the estimation of $b$.
  Interesting, the avian BMR exponent values are close to the lowest value of the predicted
  range.

  Since the heart rate scaling is obtained by using that
  $F\propto M^{-f}\propto 1/\tau\propto M^{-1/(3+x)}$,
  the predicted range for $f$ is $[-0.333,-0.286]$. There is not a recent comprehensive
  analyze of this exponent for mammals. In Tab. \ref{tab1} we also shown the data for
  the pulse rate and respiration rate of mammals and birds measured in basal conditions.
  Savage et al. \cite{savage-04} obtained $-0.25$ by binning the data of
  Brody \cite{brody-45} (original exponent $b=-0.27$). Stahl \cite{sthal-67} claims that $f=-0.25$
  but he not published its data. The empirical value for birds is more scattered and
  there is a report of an empirical value near the lower value of the predicted range.
   Although the pulse rate
  is more easy to measure than BMR, it is hard to achieve any conclusion about the empirical value of $f$.
  Note that the value $-0.27$ is out of the predicted range but is also different from $-0.25$. In fact, it
  is equidistant from $-0.25$ and $-0.286$. It is worth mentioning that $-0.27$ was also predicted by other
  recent analysis of Bishop \cite{bishop-97}. On the other hand, the respiration rate empirical
  exponents of mammals and birds have the majority of values within the predicted range.

\section{\label{sec5}Other exponents of basal metabolism for mammals and birds}

 In order to obtain other exponents, let us now characterize the network by ``aorta'' and ``capillaries''.
  We define $L_a$, $R_a$ and $v_a$ as
  the aorta length, radius and fluid velocity, respectively. The capillaries can be described by
  the capillary number $N_c$, length $l_c$, radius $r_c$ and fluid velocity $v_c$.
  It is worth mentioning that the length $l_c$ and the radius $r_c$ of capillaries
  {\sl are not necessarily invariants}, although, from our third assumption, we need some dynamical
  scaling-invariant quantities, like the blood flow speed velocity $v_0$ in the aorta or in the capillaries.
  The exponents related to these quantities can be obtained from the nutrient fluid conservation
  in the transportation network. Fluid conservation implies that
\begin{equation}
{\dot Q}=\pi R_a^2v_a=N_c\pi r_c^2v_c~~, \label{eq4}
\end{equation}
 where ${\dot Q}$ is the volume rate flow. It is clear that ${\dot Q}\propto B$ is a natural assumption.
 Since $v_a$ is invariant, the aorta radius scaling is given by $R_a\propto B^{1/2}$ and the aorta
 length scaling is described by $L_a\propto v_a\tau$. These last relationships imply that the exponents
 $a_R$ and $a_L$ defined by $R_a\propto M^{a_R}$ and $L_a\propto M^{a_L}$ are
 $a_R=(2+x)/(6+2x)$ and $a_L=1/(3+x)$. The transition from the largest length scale (aorta) to
 the cell length scale occurs in the arterioles and capillaries. $v_c$ must be scaling-invariant and,
 since the blood cells have the same size,
 we can make the extra assumption that $r_c$ is also scaling-invariant.  Therefore the density of
 capillaries $N_c/M$ behaves as $N_c/M\propto B/M$.
 The  capillary length can be invariant or mass dependent.  Since the typical cell transport length is not
 scaling-invariant, the capillary length should also depend on $M$, namely
 $l_c\propto v_c\tau_c\propto v_c\tau$.

\section{\label{sec6}The MMR of mammals and birds}

 The circulatory networks of endothermic animals make a transition from resting to maximum activity
 in such way that (i) the heart increases its rate and output, (ii) the arterial blood volume increases due 
 to constriction of the veins and (iii) the total flow and muscular flow increase, with all muscular 
 capillaries activated. These facts suggest that we have a ``forced movement'' during the characteristic time 
 $\tau$, implying that the typical constant velocity can be written as  $v=a_0\tau$
  ($a_0$ is a scaling-invariant acceleration).
  Therefore the aerobic sustained MMR is limited by  an inertial movement accelerated during time 
  $\tau$ and the ballistic movement of BMR is now given by $L=v\tau=a_0\tau^2$.

  In the upper limit of  the MMR of animals, the MMR-1 scenario, all lengths are related to the
  inertial accelerated movement ($L_i=a_i\tau^2$, $i=1,2,3$).  
  Since $V_3\propto L^3$ and $L\propto \tau^2$,
  we obtain from Eqs. (\ref{eq1}) and (\ref{eq2}) the metabolic relations:
\begin{eqnarray*}
L &\propto& M^\frac{2}{7}~~,\\
\tau&\propto& M^\frac{1}{7}~~,\\
B &\propto&  M^\frac{6}{7}~~.
\end{eqnarray*}
  This results agree with the ones obtained trough a generalization of West, Brown and Enquist \cite{west-97}
   ideas to the MMR \cite{barbosa-06}.

 In the BMR description, we had  i) a length ($L_1$) related to diffusion of $O_2$ and
 small molecules, ii) a length ($L_2$) associated to the anomalous diffusion of vesicles and
 very near to a convection movement and iii) a length ($L_3$) related the large scale transport
 of blood by convection. The MMR-1 scenario consider that $L_3$ is the only relevant length and
 that $L_1$ and $L_2$ evolved to match it. Although this description explains better the empirical
 data and is more consistent with the maximal restrictions of MMR conditions, we must consider
 other cases. We call the MMR-2 scenario, when at least the lengths related to ballistic movement
 ($L_3$ and $L_2$) changes to $L_3= a_3 \tau^2$ and $L_2=a_2\tau^2$, respectively.
 $L_1=D_1\tau^{1/2}$ remains the same.  Using Eqs. (\ref{eq1}) and (\ref{eq2}) we obtain that
 \begin{eqnarray*}
L_3&\propto& L_2 \propto M^\frac{4}{11}~~,\\
L_1&\propto& M^\frac{1}{11}~~,\\
\tau&\propto& M^\frac{2}{11}~~,\\
B &\propto&  M^\frac{9}{11}~~.
\end{eqnarray*}

 When only the length related to the large scale transports changes ($L_3=a_3\tau^2$, $L_2=v_2\tau$,
  $L_1=D_1\tau^{1/2}$), we have the MMR-3 scenario. A similar procedure furnishes the following
  results:
 \begin{eqnarray*}
L_3&\propto&  M^\frac{4}{9}~~,\\
L_2&\propto& M^\frac{2}{9}~~,\\
L_1&\propto& M^\frac{1}{9}~~,\\
\tau&\propto& M^\frac{2}{9}~~,\\
B &\propto&  M^\frac{7}{9}~~.
\end{eqnarray*}

 Note that the MMR and the heart frequency exponents should be in the intervals $[7/9,6/7]$ and
 $[-2/9,-1/7]$, respectively.
 The predicted values of $b$ and the heart rate exponent $f$ for animals agree
 with the empirical values (see Tab. \ref{tab2}) for animals in exercise-induced MMR
 conditions. But we must emphasize that the  MMR data base is much narrower than it appears.
 Several references, for example Savage et al. \cite{savage-04} and Bishop \cite{bishop-99},
  represent  basically the same data, namely those from the study of Taylor and Weibel \cite{taylor-81}
  with some variation in the data composition.
 Note also  that athletic species have a higher level of MMR than
 normal (non-athletic) ones \cite{weibel-04a}. For species with similar body mass, the MMR of athletic
 species can be 2.5 up to 5 times greater than the normal one. This implies that $\rho_4/\sigma_4$ are
 different for the two groups. The MMR theory just developed must be valid for normal species. Since the
 MMR exponent for athletic species ($b=0.94(2)$) is very different from the one ($b=0.85(2)$) for normal
 species, it is reasonable to assume that the inertial transport is different for the athletic group.
  If we assume that $L_3= c_3 \tau^3$, instead of $L_3=a\tau^2$, we obtain a large exponent
  $b=9/10$, a value near the empirical result.

 MMR can also be induced by exposure to low temperature. Oxygen consumption is measured during
 progressive reduction of the ambient temperature until a decline in this consumption
 is observed. In these experiments, the animal loses such heat quantity that the usual ways
 to dissipate it are overwhelm. Then it is possible that the relevant lengths be dominate
 by heat diffusion ($L_1\propto L_2\propto L_3\propto \tau^{1/2}$). This implies that
 $b=3/5$. However, if we consider that the blood transport in the arterial system be
 also relevant we have that $L_3=a_0t_3^2$ and $L_1\propto L_2\propto t^{1/2}$. In this case
 we obtain that $b=3/4$. It follows that the cold-induced MMR exponent should be in the interval
 $[0.600,0.750]$, in a relatively good agreement with empirical values.

\section{\label{sec7}Other exponents of maximum metabolism for mammals and birds}

The exponents related to the aorta are easily obtained from Eq. (\ref{eq4}). Now the aorta
 blood velocity is not constant and grows with body mass as  $v_a\propto a_0\tau\propto M^{1/7}$ (MMR-3).
  In fact this exponent should be in the range $[0.143,0.22]$. From now on we will discuss the exponents
  always in MMR-1 scenario which is in better agreement with the empirical data.
 The exponents
 related to the aorta radius and length are given by $a_R=5/14$ and $a_L=2/7$, respectively.
 The description of capillary scaling is not so clear. The radius $r_c$ can be assumed invariant
 since the blood cells do not depend on body mass. There are three possibilities for
 the  blood velocity $v_c$ and the capillary length $l_c$: (i) $v_c$ and $l_c$ are invariant,
 (ii) $v_c$ is invariant and $l_c=v_c\tau_c$ and (iii) $v_c=a_c\tau_c$ and $l_c=a_c\tau_c^2$ with
 $a_c$ invariant. Since we are assuming that the typical transport length in a cell is not invariant,
 $l_c$ should not be scaling-invariant. Let us consider the case (ii). Using that $r_c$ and $v_c$
 are invariant, we obtain from Eq. (\ref{eq4}) that
 the density of capillaries behaves as $N_c/M\propto M^{-1/7}$. If $\tau_c\propto\tau$ we obtain
 $l_c\propto M^{1/7}$. On the other hand, if $\tau_c\propto t_2$  and $x=1/2$ we obtain
 that $l_c\propto M^{2/7}$.

  We discuss now the aorta scaling. As already discussed , it is probable that $x=1/2$.
 In this case the BMR and MMR theories predict the same exponents $a_R$ and $a_L$ for the aorta radius
 and length, namely $a_R=0.357$ and $a_L=0.286$.
 These values are in agreement with the empirical values (see Tab.\ref{tab2}).
 The aorta velocity $v_a$ is invariant in the BMR description,
 in agreement with data (Dawson, 2003) ($v_a\propto M^{0.07}$). On the other hand, in the MMR
 description $v_a$ must depend on body mass ($v_a\propto M^{1/7}$ (MMR-1)).
 We do not know any empirical result
 for $v_a$ in the MMR conditions. So,
 it could be interesting to experimentally verify this simple prediction for normal species.

 The extra assumption that the capillaries radius is invariant is agreement with the empirical
 data \cite{schmidt-84} and with the theoretical-empirical estimations of Dawson \cite{dawson-03}
  ($r_c\propto M^{0.08}$, $l_c\propto M^{0.21}$ ). In the BMR description, the capillary velocity
  is also invariant.
  However,  the capillary length $l_c$ should  depend on $M$. If $x=1/2$ we have that
  $l_c\propto M^{0.286}$.
  In the MMR description, we must also  have $r_c$ invariant. If we assume that $v_c$ is invariant,
  we obtain that the capillary density $N_c/M\propto M^{-0.143}$ agrees well with data of
  muscular capillary density of mammals (see Tab. \ref{tab2}).
  In this case we have that $l_c=v_c\tau_c$, with
  possibilities for $\tau_c$: $\tau_c\propto\tau$ or $\tau_c\propto t_2$. Only the second
  possibility, together with $x=1/2$ is consistent with $v_c$ invariant.
  We obtain a result $l_c\propto M^{0.286}$ which agrees with the BMR description and
  with Dawson estimation.

\section{\label{sec3}The BMR of unicellular organisms}

We present now the BMR of unicellular organisms.
 The first biological length $L_1$ is related to the transport of oxygen and small molecules by
diffusion, namely $L_1=D\tau^{1/2}$. The second length is related to
 large molecules that can  be trapped  in  vesicles by
 macropinocytoses and pinocytoses and transported in direction of the nucleus.
  It is described by $L_2= D_x\tau^{(1/2)+x}$, with $x=0$ and $x=1/2$ corresponding to
  diffusion and ballistic transport, respectively.
  In order to evaluate the $b$ exponent of unicellular organisms we need to taken
 in account a third length.
  A general description of this length is achieved by supposing that
  $L_3=D_yt_3^{(1/2)+y}$,
  with $y$ varying in the interval $[0,1/2]$ (diffusion movement: $y=0$, ballistic one $y=1/2$).
 To obtain the exponent $b$ we first evaluate $V_3$ in terms of $\tau$, namely
  $V_3\propto DD_xD_y\tau^{(3/2)+x+y}$. Then, we use the relation between mass and $\tau$
  (see Eq. (\ref{eq1}) to find how $\tau$ depends on $M$ ($\tau\propto M^{2/(5+2x+2y)}$). Finally,
  we evaluate each length in terms of $M$ and
  we use  Eq. (\ref{eq2})  to obtain how $B$ depends on $M$. It follows that
  $B\propto M^{[3+2x+2y/(5+2x+2y)]}$ and $\tau\propto  M^{[2/(5+2x+2y)]}$.
  The case with only diffusion transport ($x=0$ and $y=0$)
  give-us $b=3/5$, a lower bound for the allometric exponent. The upper
  value for $x$ is $x=1/2$ (ballistic movement). In this case we have that $b=(2+y)/(3+y)$, implying
  that the allometric exponent of unicellular organisms is in the interval $[0.667,0.714]$.

   In Tab.\ref{tab3}  it is shown the empirical values of $b$ as well the predict values.
   The  value of the exponent obtained by Hemmingsen \cite{hemmingsen-60}
   ($b=0.75$) is out of our predicted range. However, Prothero \cite{prothero-86}
  observed that Hemmingsen had lumped together two metabolically different groups
 (prokaryotes and eukaryotes). When he excluded bacteria, flagellates, and marine zygotes
 from Hemmingsen's data sample, he obtained $b=0.608\pm 0.05$ for eukaryotes, a value just above
 our lower bound.
  On the other hand, Phillipson \cite{phillipson-81} studied the BMR scaling of 21 unicellular
  species and found $b=0.66\pm 0.092$.

\section{\label{sec8} Summary}

 We developed a theory for the allometric scaling of metabolism based in four
 ad-hoc postulates: i) mass density $\rho_{d+1}$ and ii) available  energy density $\sigma_{d+1}$
 are scaling-invariant quantities, iii) dominant transport processes, which are characterized
 by scaling-invariant quantities, drive the metabolic scaling
 and iv) the resource rates of these processes are matched in order to have an optimal
 nutrient delivery.

 A lower bound for all metabolic exponents, namely $b_{min}=3/5$, is found when we consider
 all transport processes as diffusive ones.

 The BMR of mammals and birds is obtained when
 we have A) diffusion, describing the transport of oxygen and small molecules, B) ballistic
 transport ($L_3\propto v_0\tau$), which is related to the blood delivery in large scale and 
 C) anomalous diffusion that represents
 the vesicular transport inside a cell and between cells of a tissue. Assuming that the last process is very
 close to the ballistic transport we obtained that $b=5/7$. This value is in good agreement with
 the best empirical estimation for BMR ($0.69$), obtained by White and Seymour (2005) for mammals
  without ruminants.
  We believe that these large mammals should be included in some way in the empirical analysis,
  implying that the real empirical $b$ value for mammals should be around $0.71$.
  The 2/3-law is obtained when the anomalous diffusion process is near normal diffusion. Therefore
  $b=2/3$ is a lower bound for BMR of mammals and birds.
  On the other hand, the 3/4-law appears as an upper bound for BMR since this value is obtained when
  all transport processes are ballistic. Therefore the $b$ exponent for mammals and birds is
   the interval $[2/3,~3/4]$. Interesting, the empirical value for birds is close to the low
   limit. The predict interval fort the exponent related to heart rate (or respiration rate) is
   $[-1/3,-1/4]$ and the most probable value is $-2/7$. They are in agreement with the empirical
   values for mammals and birds. However, this exponent was not recently studied in a comprehensive
   way as was done with the $b$ exponent.

   The aerobic sustained MMR is described by an inertial movement accelerated during time $\tau$
   ($L_3\propto a_0\tau^2$). The upper limit for the MMR exponent ($b=6/7$) is obtained when all
    transport processes are proportional to the accelerated one and the lower limit ($b=7/9$)
     occurs when only the length related to the large scale transport ($L_3$) changes to the
     accelerated movement. Since during strenuous exercise the transport systems are
   stressed to their uttermost, we believe that the upper limit describe better the MMR scaling.
   The predict value for $b$ and for the heart rate exponent are in good agreement with
   the empirical values. However, the data base is still narrow.
   The cold-induced MMR
   is studied by considering that usual heat transport processes are overwhelm and that
   we have a new metabolic state where only heat diffusion is important.
   The  different empirical exercise induced MMR exponents obtained for athletic species and
   non-athletic one can be explained qualitatively by assuming that the accelerated movement for
   athletic species is different ($L_3\propto c_0\tau^3$, for example).

   The exponents related to the aorta and capillaries of mammals
    are obtained through fluid conservation.
   Aorta blood velocity is scaling-invariant in BMR conditions but grows with mass in  the
   exercise-induced MMR situation. This  exponent, which is predicted  to be in the
   interval $[0.143,~0.22]$, was never measured. The empirical determination of this exponent
   seems to be easy and interesting. Moreover, it can be a experimental test of the importance
   of the transportation processes for the metabolic scaling.
   The exponents characterizing the length and radius of aorta and capillaries have the same
   values in the BMR description and in the upper limit of exercise-induced MMR situation.
   The predict values agree with the empirical ones. On the other hand, the capillary density
   must be described by the MMR scenario and the predict value also agrees with the experimental
   value.

   Finally we discussed the BMR of unicellular organisms. In this case two transport processes
   are diffusion, related to oxygen and small molecules transport, and anomalous diffusion,
   related to vesicular transport. The third transport process was described also as
   an anomalous diffusion one because it allow us to change the transport mechanism
    from diffusion until a
   ballistic movement. The predicted range $[2/3,~5/7]$ was  compared with the few
   results of the literature.

\noindent{ $^{\ast}$      Electronic address: jaff@fisica.ufmg.br\\}
$^{\dagger}$   Electronic address: lau@ifsc.usp.br\\

\section{Acknowledgements}

 JKL thanks CNPq, CAPES and FAPEMIG for financial support.  LAB thanks FAPESP for financial support.


\newpage

\begingroup
\squeezetable
\begin{table}
\caption{Allometric exponents related to the basal metabolism for mammals and birds.
Under parenthesis is the error in the last significative of the observed quantities.
 \label{tab1} }
\begin{center}
\begin{tabular}{|c|c|c|c|}
\hline
     Predicted          &      Observed & Comment      & Ref. \\
\hline
\multicolumn{4}{|c|} {Basal metabolic rate}        \\
\hline
      $0.714$           & $0.712(13)$ & mammals   & \cite{savage-04} \\
      $[0.667,0.714]$   & $0.737(26)$ & binning data & \cite{savage-04} \\
      $0.666$ (BMR-2)   & $0.668(25)$ & small mammals & \cite{dodds-01}   \\
      $0.750$ (BMR-1)   & $0.710(21)$ & mammals   & \cite{dodds-01}   \\
                        & $0.664(15)$ & birds       & \cite{dodds-01}  \\
                        & $0.686(14)$ & large mammals excluded & \cite{white-05} \\
                        & $0.669$     & birds       &\cite{mckechnie-04}   \\
\hline
\multicolumn{4}{|c|} {Heart rate}        \\
\hline
 $-0.286$              &  $-0.25(2)$ & mammals, unknown data  & \cite{sthal-67} \\
 $[-0.333,-0.286]$     &  $-0.25$   & mammals               & \cite{gunther-75}  \\
 $-0.333$ (BMR-2)      &  $-0.27$  & mammals & \cite{brody-45,adolph-43,gunther-66a}    \\
 $-0.250$ (BMR-1)      &  $-0.25(3)$ & mammals, binning data of \cite{brody-45}& \cite{savage-04}   \\
                       &  $-0.26$ & mammals           & \cite{weibel-04b}   \\
                       &  $-0.27$ & mammals           & \cite{li-00}   \\
                       &  $-0.23$ & birds      & \cite{lasiewiski-71}   \\
                       &  $-0.209$ & birds       & \cite{berger-74}   \\
                       &  $-0.33(6)$ & birds       & \cite{bishop-95}  \\
\hline
\multicolumn{4}{|c|} {Respiration rate}\\
\hline
 $-0.286$              &  $-0.260(5)$ & mammals  & \cite{sthal-67}  \\
 $[-0.333,-0.286]$     &  $-0.28$   & mammals    & \cite{gunther-66a} \\
 $-0.333$ (BMR-2)    &  $-0.28$  & mammals & \cite{adolph-49} \\
 $-0.250$ (BMR-1)    &  $-0.25$  & mammals & \cite{guyton-47}\\
                     &  $-0.26(6)$ & mammals, binning data of \cite{calder-68}& \cite{savage-04}  \\
                     &  $-0.31$ & birds      & \cite{lasiewiski-71}   \\
                     &  $-0.20$ & birds       & \cite{berger-74} \\
                     &  $-0.33$ & passerines   & \cite{gunther-75}   \\
                     &  $-0.28$ & nonpasserines & \cite{gunther-75}   \\
\hline
\end{tabular}
\end{center}
\end{table}
\endgroup

\begin{table}
\caption{Allometric exponents related to maximum metabolism and other exponents for mammals and birds.
Under parenthesis is the error in the last significative of the observed quantities.
 \label{tab2} }
\begin{center}
\begin{tabular}{|c|c|c|c|}
\hline
\multicolumn{4}{|c|} {Exercise-induced maximum metabolic rate}\\
\hline
   $0.857$ (MMR-1)      &  $0.828(70)$ & mammals, binning data  & \cite{savage-04}    \\
 $[0.778,0.857]$        &  $0.88(2)$   & standard animals    & \cite{bishop-99}   \\
                        &  $0.872(29)$ & mammals          & \cite{weibel-04a}    \\
                        &  $0.94(2) $  & athletic   & \cite{weibel-04a}     \\
                        &  $0.85(2)$ & non-athletic    & \cite{weibel-04a} \\
                        &  $0.882(24)$ & marsupials    & \cite{hinds-93}   \\
                        &  $0.87(5)$ & mammals          & \cite{white-05}   \\
\hline
\multicolumn{4}{|c|} {Heart rate}\\
\hline
 $-0.143$ (MMR-1)       &  $-0.16(2)$  & mammals & \cite{bishop-97}\\
$[-0.222,-0.143]$       &  $-0.15$ & mammals     &\cite{weibel-04b}   \\
                        &  $-0.17(2)$ & birds, flight &\cite{bishop-95}    \\
                        &  $-0.146$ & birds, flight & \cite{berger-74}   \\
                        &  $-0.157$ & birds            & \cite{berger-74}    \\
\hline
\multicolumn{4}{|c|} {Cold-induced maximum metabolic rate}\\
\hline
 $[0.600,0.750]$        &  $0.65(5)$ & mammals           & \cite{white-05}   \\
                        &  $0.772(30)$ & marsupials   &  \cite{hinds-93}   \\
                        &  $0.789(40)$ & theria   & \cite{hinds-93}  \\
\hline
\multicolumn{4}{|c|} {Capillary density}\\
\hline
 $-0.143$ (MMR-1)         &  $-0.14(7)$   &      & \cite{calder-84,hoppeler-81} \\
\hline
\multicolumn{4}{|c|} {Aorta radius }\\
\hline
 $0.357$ (BMR and MMR-1)  &  $0.36$ & mammals       &\cite{calder-84,li-00} \\
 $[0.333, 0.357]$ (BMR)   &  $0.335$ & mammals & \cite{gunther-66a}   \\
\hline
\multicolumn{4}{|c|} {Aorta length }\\
\hline
 $0.286$ (BMR and MMR-1) &  $0.32$ & mammals  & \cite{gunther-66a,calder-84}   \\
 $[0.286, 0.333]$ (BMR) &  $0.31$ & mammals      & \cite{li-00} \\
\hline
\end{tabular}
\end{center}
\end{table}

\begin{table}
\caption{ BMR allometric exponent $b$
 ($B\sim M^b$) for unicellular organisms.
Under parenthesis is the error in the last significative of the observed quantities.
 \label{tab3} }
\begin{center}
\begin{tabular}{|c|c|c|}
\hline
 \multicolumn{3}{|c|} { Exponent $b$ }       \\
\hline
                 Predicted            & Observed           & Ref. \\
\hline
                 $0.714$              &  $0.75$        & \cite{hemmingsen-60}\\
                 $[0.667, 0.714]$     &  $0.66(9)$     & \cite{phillipson-81}\\
                 $0.600$ (MR-3)      &  $0.608(5)$    & \cite{prothero-86}\\
\hline

\end{tabular}
\end{center}
\end{table}

\end{document}